\documentclass[twocolumn,aps,prl,superscriptaddress,floatfix]{revtex4-1}

\usepackage[version=3]{mhchem}
\usepackage{graphicx}
\newcommand{\forceindent}{\leavevmode{\parindent=1em\indent}}

\begin{document}
	\title{Abnormal $phase$ flip in coherent phonon oscillations of Ca$_2$RuO$_4$}
		
	\author{Min-Cheol Lee}
	\author{Choong H. Kim}
	\author{Inho Kwak}
	\affiliation{Center for Correlated Electron Systems (CCES), Institute for Basic Science (IBS), Seoul 08826, Republic of Korea}
	\affiliation{Department of Physics and Astronomy, Seoul National University, Seoul 08826, Republic of Korea}
	\author{J. Kim}
	\author{S. Yoon}
	\affiliation{Department of Physics and New Renewable Energy Research Center (NREC), Ewha Womans University, Seoul 03760, Republic of Korea}
	\author{Byung Cheol Park}
	\author{Bumjoo Lee}
	\affiliation{Center for Correlated Electron Systems (CCES), Institute for Basic Science (IBS), Seoul 08826, Republic of Korea}
	\affiliation{Department of Physics and Astronomy, Seoul National University, Seoul 08826, Republic of Korea}
	\author{F. Nakamura}
	\affiliation{Department of Education and Creation Engineering, Kurume Institute of Technology, Fukuoka 830-0052, Japan}
	\author{C. Sow}
	\author{Y. Maeno}
	\affiliation{Department of Physics, Graduate School of Science, Kyoto University, Kyoto 606-8502, Japan}
	\author{T. W. Noh}
	\thanks{Corresponding author}
	\email{twnoh@snu.ac.kr}
	\affiliation{Center for Correlated Electron Systems (CCES), Institute for Basic Science (IBS), Seoul 08826, Republic of Korea}
	\affiliation{Department of Physics and Astronomy, Seoul National University, Seoul 08826, Republic of Korea}
	\author{K. W. Kim}
	\thanks{Corresponding author}
	\email{kyungwan.kim@gmail.com}
	\affiliation{Department of Physics, Chungbuk National University, Cheongju, Chungbuk 28644, Republic of Korea}

	\date{\today}

\begin{abstract}
We employ an optical pump-probe technique to study coherent phonon oscillations in Ca$_2$RuO$_4$. We find that oscillation-amplitude of an $A_g$ symmetric phonon mode is strongly suppressed at 260 K, a putative transition point of orbital ordering. The oscillation also shows a gradual but huge change in its $phase$, such that the oscillation even flips over with a 180$^{\circ}$ change across the temperature. Density functional theory calculations indicate that the $A_g$ phonon has an eigenmode of octahedral distortion with conventional tilting along the $a$-axis and antipolar distortion of apical oxygen. Careful inspection of the lattice captures an unusually large antipolar distortion in low-temperature structures, which may play a crucial role for the phase transition at 260 K. 
\end{abstract}
\maketitle
Ca$_2$RuO$_4$ is a prototype Mott insulator, where all of the degrees of freedom — charge, spin, orbital and lattice — show robust interactions in distinctive phase transitions \cite{Nakatsuji1997, Braden1998, Friedt2001,Jain2017,Lee2017,Zegkinoglou2005,Porter2018}. A metal-insulator transition (MIT) occurs at $T_\textnormal{MIT}$ = 357 K, accompanied by a structural transition that involves strong distortions of octahedral flattening and tilting \cite{Nakatsuji1997,Friedt2001}. Upon cooling, antiferromagnetic (AFM) spin ordering develops in Ca$_2$RuO$_4$ below $T_\textnormal{N}$ = 113 K, where its magnetism is determined by the spin-orbit coupling and the tetragonal distortions \cite{Braden1998,Jain2017}. Recently, the sizable spin-phonon coupling has been observed in coherent phonon oscillations \cite{Lee2017}. Another interesting anomaly has been observed below $T_\textnormal{OO}$ = 260 K, where an orbital ordering has been suggested. Resonant x-ray scattering (RXS) experiments indicated that the antiferromagnetic diffraction peak shows up even in the paramagnetic phase below $T_\textnormal{OO}$ \cite{Zegkinoglou2005,Porter2018}. However, the exact configuration and origin of the ordering remain unresolved. Apart from these RXS results, there have been no further experimental observations on the 260 K anomaly, nor has any theoretical model been presented to support the order.\\
\forceindent Recent advances in ultrafast techniques have rendered it possible to investigate novel phenomena in non-equilibrium states \cite{Huber2001, Brorson1990, Mankowsky2014, Kubler2007, Kim2012,Gerber2015,Schmitt2008,Schafer2010,Forst2011,Polli2007,Dhar1994}. Of particular interest is coherent phonon oscillations, which result in oscillations of a probing signal arising from periodic modulation of the lattice potential \cite{Dhar1994,Zeiger1992,Stevens2002,Kuznetsov1994}. In contrast to thermally activated phonons with random \textit{phase}s, coherent phonons provide an oscillation-\textit{phase} value that reflects physical properties of a material. However, this \textit{phase} tends to be overlooked, as it has been believed to be simply determined by the generation mechanism \cite{Dhar1994,Zeiger1992,Stevens2002,Kuznetsov1994}.\\
\forceindent In this study, we find that the coherent phonon oscillations of Ca$_2$RuO$_4$ exhibit huge anomalies across $T_\textnormal{OO}$. To our surprise, one $A_g$ phonon mode changes its oscillation-$phase$ showing an unexpected 180$^{\circ}$ flip. Density functional theory (DFT) calculations find that the $A_g$ phonon mode is of octahedral tilting vibrations that are nearly parallel to the structural deformation at $T < T_\textnormal{OO}$. Scrutiny of the temperature dependent octahedral structure reveals that a lattice deformation with a large antipolar distortion along the $b$-axis develops below $T_\textnormal{OO}$.\\
\forceindent We perform time-resolved reflectance measurements on single-crystalline Ca$_2$RuO$_4$, synthesized by the floating zone method \cite{Nakatsuji2001}. We utilize near infrared 800-nm pulses for both the pump and probe beams generated from a commercial Ti:Sapphire amplifier system with a 250-kHz repetition rate. The corresponding photon energy of 1.55 eV is much greater than the optical gap of 0.6 eV of Ca$_2$RuO$_4$ \cite{Jung2002}. The time duration of the pump and probe pulses are 30 fs. The oscillations show a linear response to pump fluence over a wide fluence range up to 1 mJ/cm\ce{^2} (Fig. S1). We present data measured at pump and probe fluences of 140 and 80 \ce{$\mu$}J/cm\ce{^2}, respectively, to minimize the heating effect and to maintain the measurement conditions close to the linear response region of the electronic response. The full width half maximum (FWHM) spot sizes of the pump and probe pulses were 80 and 40 $\mu$m, respectively. The pump and probe pulses are linearly polarized and perpendicular to each other. We cannot find a noticeable anisotropy of the response depending on both pump and probe polarizations.\\
\begin{figure}[!t]
	\includegraphics[width=3.5 in]{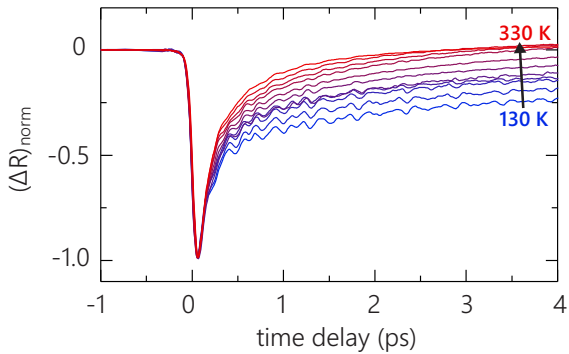}
	\caption{(color online) Photo-induced reflectivity change normalized by maximum peaks for clarity. The photo-induced reflectivity transient is measured at every 20 K from 130 to 330 K.}
	\label{Figure1}
\end{figure}
\begin{figure*}[]
	\includegraphics[width = 7 in]{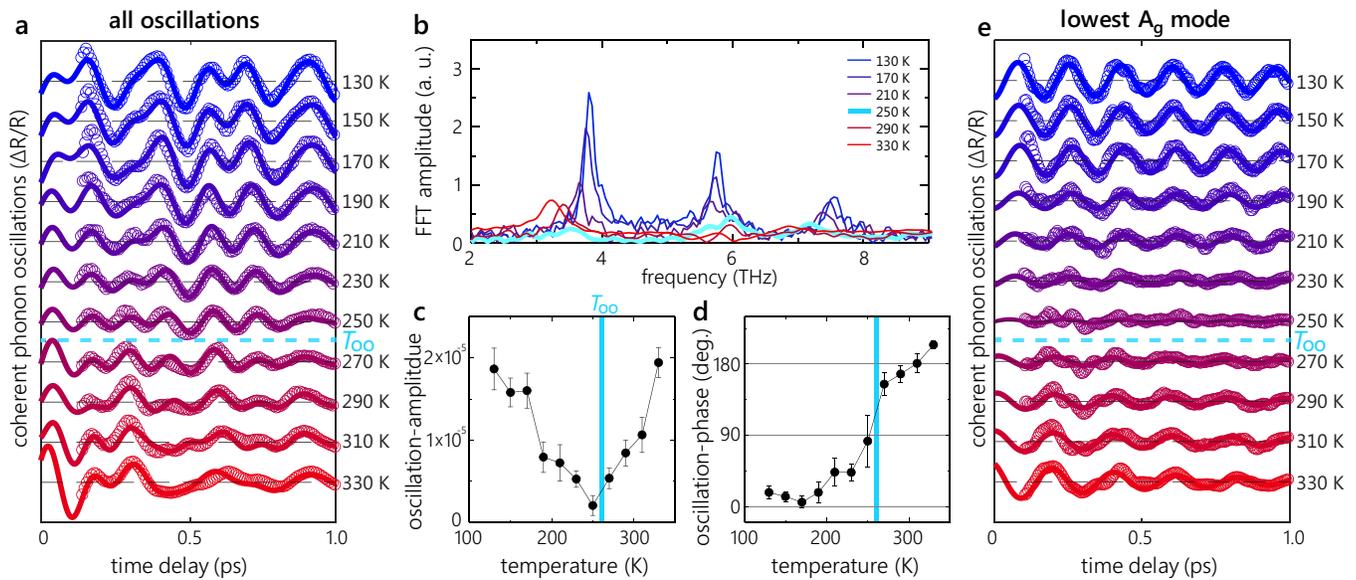}
	\caption{(color online) (a) Coherent phonon oscillations of photo-induced reflectivity change (open circles). The data have been subtracted using a bi-exponential decay fitting (solid lines) at temperature intervals of 20 K from 130 to 330 K. These data are not normalized by maximum peaks. (b) Fourier transformation data of the oscillating components in (a). All oscillating components correspond to $A_g$ symmetric Raman modes. (c,d) $T$-dependent fitting parameters for the amplitude and $phase$ of the lowest-frequency mode. (e) The oscillating component (open circles) and fit curves (solid lines) for the lowest-frequency mode. Higher-frequency oscillations have been subtracted from the raw data using the damped harmonic oscillator model fitting. The lowest $A_g$ oscillation exhibits clear anomalies across $T_\textnormal{OO}$ = 260 K, where the amplitude is nearly suppressed and the oscillation-\textit{phase} changes with a 180$^{\circ}$ flip.}
	\label{Figure2}
\end{figure*}
\forceindent Figure 1 shows the photoinduced reflectivity change of Ca$_2$RuO$_4$ after near-infrared pumping at various temperatures from 130 to 330 K. The data are normalized by the maximum peak values at each temperature for a clear comparison. Analysis using a bi-exponential function fitting (Fig. S2) indicates that relaxation processes exhibit two decay time scales of 0.1 ps and 1 ps. The electronic responses of both relaxation times do not show a noticeable anomaly across $T_\textnormal{OO}$.\\
\forceindent Superimposed over the overall relaxation, periodic oscillations in the transient reflectivity are clearly observed. The oscillating components obtained by subtracting the electronic responses by means of the bi-exponential curve fitting are shown in Fig. 2(a). Fourier transform analysis reveals that the coherent oscillations are composed of multiple phonon modes as shown in Fig. 2(b). All of the modes correspond to $A_g$ symmetric phonons that have been observed in the previous Raman experiments \cite{Rho2005,Souliou2017}. Interestingly, the $A_g$ phonon mode of the lowest frequency of 3.8 THz gets almost fully suppressed at $T \sim T_\textnormal{OO}$ = 260 K. The suppression also appears in Raman scattering spectra as shown in Fig. S3, although previous Raman studies did not focus on the transition at $T_\textnormal{OO}$ \cite{Rho2005,Souliou2017}.\\
\forceindent To extract quantitative information from the coherent oscillations, we fit the data with a damped harmonic oscillator model: $\Delta{R}_{CP}(t)$= $-\Sigma_i{A_i}\cos(2\pi{f}_i+\phi_i)\exp(t/\tau_i)$, where ${A_i}, {f_i}, {\phi_i}$, and ${\tau_i}$ present the amplitude, frequency, initial \textit{phase}, and damping time of the $A_g$ symmetric modes, respectively. The fitting results are shown as line curves in Fig. 2(a); all of the curves well-matched to the measurement data. Figure 2(c) and 2(d) show the oscillation-amplitude and $phase$ of the 3.8 THz mode, respectively. Both parameters show clear anomalies, i.e., not only suppression of the amplitude, but also huge variation in the phase across $T_\textnormal{OO}$. These anomalies show up also in the time domain signal of the $A_g$ phonon oscillations of the lowest frequency mode after the subtraction of higher-frequency components above 4 THz, as shown in Fig. 2(e). Such a phase variation by 180$^{\circ}$ of flipping is unexpected without a structural transition. As far as we know, the oscillation-$phase$ flip has been reported previously only once in blue bronze across a structural phase transition \cite{Schafer2010}.\\
\begin{figure}[!b]
	\includegraphics[width = 3.5 in]{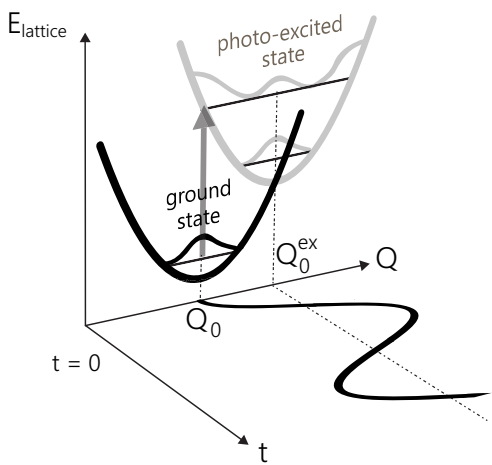}
	\caption{(color online) Schematic diagram of conventional generation mechanism of displacive phonon oscillations. The arrow indicates the optical transition resulting from 1.55 eV pumping. Vibrational functions of $n = 0$ and $n = 1$ modes are also displayed in each lattice potential.}
	\label{Figure3}
\end{figure}
\forceindent The observed $phase$ values in Fig. 2(e) suggest that the cosine-type phonon oscillations are dominant in the paramagnetic insulating state well below and above $T_\textnormal{OO}$. Such cosine-type oscillations have been explained by the displacive-type generation \cite{Zeiger1992}. The generation process is shown schematically in Fig. 3 in terms of the lattice potential as a function of lattice coordinate ($Q$) depending on the absorption. The essence is that photo-excitation can shift the minimum position of the lattice potential from $Q_0$ to $Q_0^{ex}$, because of an instantaneous change in the charge density distribution \cite{Zeiger1992,Stevens2002,Kuznetsov1994}. Using optical ellipsometry techniques, we confirm that Ca$_2$RuO$_4$ are always strongly absorbing the pump beam at all measured temperatures (Fig. S4), which is consistent with the displacive-type oscillations. Although there is a gradual change in absorption across $T_\textnormal{OO}$, the 180$^{\circ}$ $phase$ flip is unexpected based on the case of the abovementioned displacive type coherent phonons.\\
\forceindent What is the origin of the $phase$-flip in Ca$_2$RuO$_4$? Modulations of the reflectivity due to phonon oscillations should follow $\Delta{R}_\textnormal{CP} = (\partial{R}/\partial{Q})\delta{Q}$ \cite{Zeiger1992,Stevens2002,Kuznetsov1994}. The initial displacement $\delta{Q}$ at $t = 0$ is determined by the shift of the minimum of the lattice potential, $Q_0^{ex}-Q_0$ in Fig. 3. If $\delta{Q}=Q_0^{ex}-Q_0$ gradually changes its sign, the flip of the reflectivity modulations may occur (Fig. S5). The gradual sign change of $\delta{Q}$ is also consistent with the suppression of the oscillation amplitude when $\delta{Q}\sim0$ around $T_\textnormal{OO}$. On the other hand, the \textit{phase}-flip could occur when $(\partial{R}/\partial{Q})$ changes the sign while $\delta{Q}$ stays with the same sign. The change of the reflectivity on the phonon displacement is given by $\partial R/\partial Q = (\partial R/\partial \epsilon_1)(\partial \epsilon_1 /\partial Q) + (\partial R/\partial \epsilon_2)(\partial \epsilon_2 /\partial Q)$ \cite{Zeiger1992}. The contributions of $(\partial \epsilon_1 /\partial Q)^2$ and $(\partial \epsilon_2 /\partial Q)^2$ are proportional to the Raman cross section \cite{Zeiger1992}, the intensity of which shows an anomaly across $T_\textnormal{OO}$ as illustrated in Fig. S3. Thus, it is also possible that Raman susceptibility changes sign across the ordering temperature, whereas the sign of the initial phonon displacement $\delta{Q}$ is invariant.\\
\forceindent To investigate the coupling between the phonon and the phase transition at $T_\textnormal{OO}$, we perform DFT calculations. The phonon eigenmodes are calculated by frozen phonon method. The eigenmode of the lowest $A_g$ mode is shown in Fig. 4(a) and Fig. 4(b); we display only four neighboring RuO$_6$ octahedra, omitting the motions of Ca$^{2+}$ because the cations with fully occupied electron shells hardly influence the optical responses. The $A_g$ mode can be described by octahedral tilting associated with motions of both in-plane oxygen ($O_\textnormal{P}$) and apical oxygen ($O_\textnormal{A}$). The motion of $O_\textnormal{P}$ tilts along the $a$-axis. However, we find that $O_\textnormal{A}$ vibrates along the $-x$ [$+y$] direction at the site of the Ru(1) [Ru(2)] ion. The diagonal motion of $O_\textnormal{A}$ in Ca$_2$RuO$_4$ is unusual, distinct from $O_\textnormal{A}$ motions in usual $A_g$ phonon modes of a tilting character in other layered compounds, which show rigid tilts of octahedral along the unit cell axis \cite{Cohen1988,Sohn2014}.\\
\forceindent We find that the $O_\textnormal{A}$ position of Ca$_2$RuO$_4$ qualitatively becomes different across $T_\textnormal{OO}$. We closely examined the crystal structure associated with $O_\textnormal{A}$. Figures 4(c) and 4(d) show simplified diagrams of the octahedral distortions, which are obtained from a close scrutiny of the previously reported neutron scattering experiments at 295 and 180 K \cite{Braden1998,Friedt2001}. The octahedral rotations around the $c$-axis are ignored in the diagrams for simplicity. The $O_\textnormal{A}$ positions at 295 K ($>T_\textnormal{OO}$) represent usual octahedral tilting but with a discrepancy in tilting angles $O_\textnormal{A}$ and $O_\textnormal{P}$. The $O_\textnormal{A}$ tilting angle ($\theta_\textnormal{A}$) is smaller than the $O_\textnormal{P}$ tilting angle ($\theta_\textnormal{P}$). As a result, when projected normally to the local $O_\textnormal{P}$ planes, $O_\textnormal{A}$ is shifted from the symmetric center positions in the $-a$ [$+a$] direction at the Ru(1) [Ru(2)] site, respectively, as shown in Fig. 4(c). However, the $O_\textnormal{A}$ positions at 180 K ($<T_\textnormal{OO}$) are shifted along the diagonal directions along $-x$ [$+y$] in Ru(1) [Ru(2)] octahedral as depicted in Fig. 4(d). This contrasts clearly with the 295 K structure. \\
\begin{figure}[!t]
	\includegraphics[width = 3.3 in]{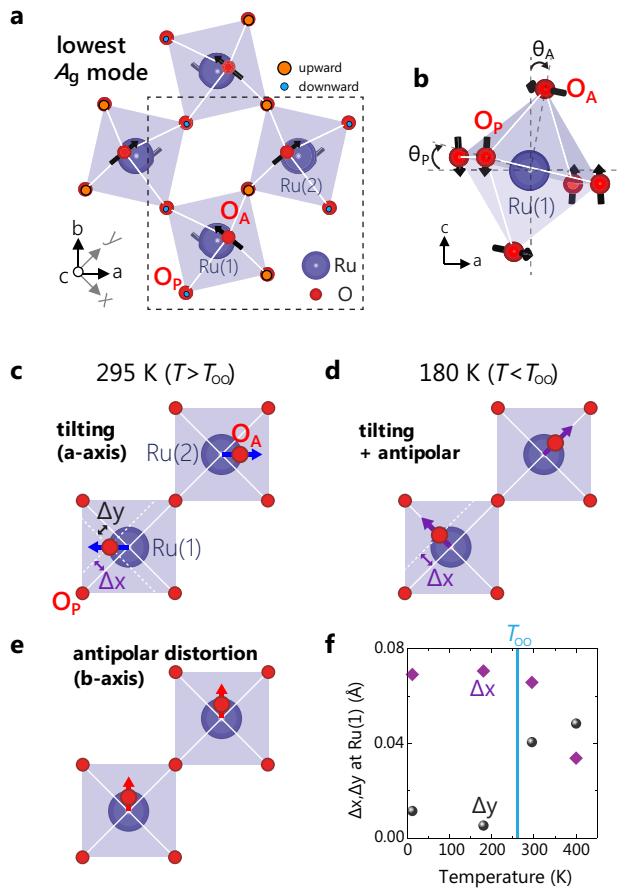}
	\caption{(color online) (a, b) Eigenmode of the lowest $A_g$ phonon mode. The black arrows indicate vibrations of the apical oxygen atoms ($O_\textnormal{A}$). The modulations of in-plane oxygen atoms ($O_\textnormal{P}$) in upward (orange circle) and downward (sky blue circle) directions have a magnitude comparable to that of $O_\textnormal{A}$. The dotted box indicates the pair of neighboring RuO$_6$ octahedra which are described in (c-e) schematically. (c, d) Simplified diagrams of the octahedral distortions (ignoring rotation) (b) above $T_\textnormal{OO}$ and (c) below $T_\textnormal{OO}$. At 295 K ($>T_\textnormal{OO}$), the difference in tilting angle between $O_\textnormal{A}$ and $O_\textnormal{P}$ causes shifts in the positions of $O_\textnormal{P}$ from the symmetric center positions. In particular, the structural distortions in (d) are nearly parallel to the eigenmode of the $A_g$ phonon mode in (a). (e) Schematic diagram of the antipolar distortion. (f) Atomic displacement of the $O_\textnormal{A}$ in Ru(1) ion. All data are extracted from the previous neutron scattering results \cite{Braden1998, Friedt2001}.}
	\label{Figure4}
\end{figure}
\forceindent The discrepancy between the 180 K and 295 K structures can be understood by an additional antipolar distortion of $O_\textnormal{A}$ toward the $b$-axis, as shown in Fig. 4(e). The 180 K structure can be obtained if the antipolar distortion is added to the structure at 295 K. This results in site-dependent local anisotropy below $T_\textnormal{OO}$. The neutron scattering data available at four temperatures \cite{Braden1998,Friedt2001} indicate that the antipolar distortion develops around $T_\textnormal{OO}$ as shown in Fig. 4(f). The relative displacement of $O_\textnormal{A}$ at the Ru(1) [Ru(2)] site is almost pure $\Delta{x}$ [$\Delta{y}$] at 180 K below $T_\textnormal{OO}$, while the $\Delta{x}$ and $\Delta{y}$ components are comparable at 295 K above $T_\textnormal{OO}$. Indeed, the tilting angle of $O_\textnormal{A}$ increases more steeply than $O_\textnormal{P}$ below 260 K \cite{Bradentilt}, which might be attributed due to the development of the antipolar distortion. We note that the $O_\textnormal{A}$ distortions from the symmetric position at 180 K are exactly parallel to the apical motions of the lowest $A_g$ phonon mode. Therefore, it is natural that the phonon oscillations are sensitive to the temperature dependent octahedral deformation. We suggest that the change of the position of the apical oxygen may result in a sign change in either of $\delta{Q}$ or $(\partial R/\partial Q)$. Verifying the exact origin of the $phase$-flip in Ca$_2$RuO$_4$ requires additional time-resolved measurement on the lattice structures, using X-ray pump–probe spectroscopy.\\
\forceindent In summary, we investigate the coherent phonon oscillations in Ca$_2$RuO$_4$ that show a 180$^\circ$ $phase$ variation across $T_\textnormal{OO}$ = 260 K. Careful inspection of the crystal structure provides an evidence of structural evolution of octahedra with the development of antipolar distortion below $T_\textnormal{OO}$. These observations put cornerstones to understand the mysterious 260 K transition in Ca$_2$RuO$_4$. Our results emphasize that $phase$-sensitive measurements of coherent oscillations offer a unique opportunity to investigate quantum phase transitions coupled to the lattice in complex materials.\\

\begin{acknowledgements}
This work was supported by the Institute for Basic Science (IBS) in Korea (IBS-R009-D1). K.W.K. was supported by the Basic Science Research Program through the National Research Foundation of Korea (NRF) funded by the Ministry of Science, ICT and Future Planning (NRF-2015R1A2A1A10056200 and 2017R1A4A1015564). J. Kim and S. Yoon were supported by NRF-2016R1D1A1B01009032. This work was supported by JSPS KAKENHI (Nos. JP15H05852, JP15K21717 and JP17H06136), JSPS Core-to-Core program. C.S acknowledges support of the JSPS International Research Fellowship (No. JP17F17027).
\end{acknowledgements}


\begin{references}
\bibitem{Nakatsuji1997}	S. Nakatsuji, S. I. Ikeda and Y. Maeno, J. Phys. Soc. Japan \textbf{66}, 1868 (1997).
\bibitem{Friedt2001}	O. Friedt, M. Braden, G. Andre, P. Adelmann, S. Nakatsuji, and Y. Maeno, Phys. Rev. B \textbf{63}, 174432 (2001).
\bibitem{Braden1998} M. Braden, G. Andre, S. Nakatsuji, and Y. Maeno, Phys. Rev. B \textbf{58}, 847 (1998).
\bibitem{Jain2017} A. Jain, M. Krautloher, J. Porras, G. H. Ryu, D. P. Chen, D. L. Abernathy, J. T. Park, A. Ivanov, J. Chaloupka, G. Khaliullin, B. Keimer, and B. J. Kim, Nat. Phys. \textbf{13}, 633 (2017).
\bibitem{Lee2017} M.-C. Lee, C. H. Kim, I. Kwak, C. W. Seo, C. H. Sohn, F. Nakamura, C. Sow, Y. Maeno, E.-A. Kim, T. W. Noh, and K. W. Kim, arXiv 1712.03028 (2017).
\bibitem{Zegkinoglou2005} I. Zegkinoglou, J. Strempfer, C. S. Nelson, J. P. Hill, J. Chakhalian, C. Bernhard, J. C. Lang, G. Srajer, H. Fukazawa, S. Nakatsuji, Y. Maeno, and B. Keimer, \textbf{95}, 136401 (2005).
\bibitem{Porter2018} D. G. Porter, V. Granata, F. Forte, S. Di Matteo, M. Cuoco, R. Fittipaldi, A. Vecchione, and A. Bombardi, arXiv 1807.00721 (2018).
\bibitem{Huber2001}	R. Huber, F. Tauser, A. Brodschelm, M. Bichler, G. Abstreiter, and A. Leitenstorfer, Nature \textbf{414}, 286 (2001).
\bibitem{Brorson1990}	S. D. Brorson, A. Kazeroonian, J. S. Moodera, D. W. Face,T. K. Cheng, E. P. Ippen, M. S. Dresselhaus, G. Dresselhaus, Phys. Rev. Lett. \textbf{64}, 2172 (1990).
\bibitem{Mankowsky2014}	R. Mankowsky, A. Subedi, M. F{\"o}rst, S. O. Mariager, M. Chollet, H. T. Lemke, J. S. Robinson, J. M. Glownia, M. P. Minitti, A. Frano, M. Fechner, N. A. Spaldin, T. Loew, B. Keimer, A. Georges, and A. Cavalleri., Nature \textbf{516}, 71 (2014).
\bibitem{Kubler2007}	C. K{\"u}bler, H. Ehrke, R. Huber, R. Lopez, A. Halabica, R. F. Haglund, and A. Leitenstorfer, Phys. Rev. Lett. \textbf{99}, 116401 (2007).
\bibitem{Kim2012}	K. W. Kim, A. Pashkin, H. Sch{\"a}fer, M. Beyer, M. Porer, T. Wolf, C. Bernhard, J. Demsar, R. Huber and A. Litenstorfer, Nat. Mater. \textbf{11}, 497 (2012).
\bibitem{Gerber2015}	S. Gerber, K. W. Kim, Y. Zhang, D. Zhu, N. Plonka, M. Yi, G. L. Dakovski, D. Leuenberger, P. S. Kirchmann, R. G. Moore, M. Chollet, J. M. Glownia, Y. Feng, J.-S. Lee, A. Mehta, A. F. Kemper, T. Wolf, Y.-D. Chuang, Z. Hussain, C.-C. Kao, B. Moritz, Z.-X. Shen, T. P. Devereaux and W.-S. Lee, Nat. Commun. \textbf{6}, 7377 (2015).
\bibitem{Schmitt2008}	F. Schmitt, P. S. Kirchmann, U. Bovensiepen, R. G. Moore, L. Rettig, M. Krenz, J.-H. Chu, N. Ru, L. Perfetti, D. H. Lu, M. Wolf, I. R. Fisher, and Z.-X. Shen, Science \textbf{321}, 1649 (2008).
\bibitem{Schafer2010}	H. Sch{\"a}fer, V. V Kabanov, M. Beyer, K. Biljakovic, and J. Demsar, Phys. Rev. Lett. \textbf{105}, 066402 (2010).
\bibitem{Forst2011}	M. F{\"o}rst, C. Manzoni, S. Kaiser, Y. Tomioka, Y. Tokura, R. Merlin, and A. Cavalleri, Nat. Phys. \textbf{7}, 854 (2011).
\bibitem{Polli2007}	D. Polli, M. Rini, S. Wall, R. W. Schoenlein, Y. Tomioka, Y. Tokura, G. Cerullo, and A. Cavalleri, Nat. Mater. \textbf{6}, 643 (2007).
\bibitem{Dhar1994}	L. Dhar, J. A. Rogers, and K. A. Nelson, Chem. Rev. \textbf{94}, 157 (1994).
\bibitem{Zeiger1992}	H. J. Zeiger, J. Vidal, T. K. Cheng, E. P. Ippen, G. Dresselhaus, and M. S. Dresselhaus, Phys. Rev. B \textbf{45}, 768 (1992).
\bibitem{Stevens2002} T. E. Stevens, J. Kuhl, and R. Merlin, Phys. Rev. B \textbf{65}, 144304 (2002).
\bibitem{Kuznetsov1994} A. V. Kuznetsov and C. J. Stanton, Phys. Rev. Lett. \textbf{73}, 3243 (1994)
\bibitem{Nakatsuji2001}	S. Nakatsuji and Y. Maeno, J. Solid State Chem. \textbf{156}, 26 (2001).
\bibitem{Jung2002} J. H. Jung, Z. Fang, J. P. He, Y. Kaneko, Y. Okimoto, and Y. Tokura, Phys. Rev. Lett. \textbf{91}, 056403 (2003).
\bibitem{Rho2005}	H. Rho, S. L. Cooper, S. Nakatsuji, H. Fukazawa, and Y. Maeno, Phys. Rev. B \textbf{71}, 245121 (2005).
\bibitem{Souliou2017}	S.-M. Souliou, J. Chaloupka, G. Khaliullin, G. Ryu, A. Jain, B. J. Kim, M. Le Tacon, and B. Keimer, Phys. Rev. Lett. \textbf{119}, 067201 (2017).
\bibitem{Cohen1988} R. E. Cohen, W. E. Pickett, H. Krakauer, and L. L. Boyer, Physica B \textbf{150}, 61 (1988).
\bibitem{Sohn2014} C. H. Sohn, M.-C. Lee, H. J. Park, K. J. Noh, H. K. Yoo, S. J. Moon, K. W. Kim, T. F. Qi, G. Cao, D.-Y. Cho, and T. W. Noh, Phys. Rev. B \textbf{90}, 041105 (2014).
\bibitem{Bradentilt} See Fig. 6 in Ref. [3] of Phys. Rev. B \textbf{58}, 847 (1998).
\end{references}
\end{document}